\documentstyle[11pt]{article}

\makeatletter
                                                                               
\def\icite{\@ifnextchar [{\@tempswatrue\@citey}{\@tempswafalse\@citey[]}}
\def\@citex[#1]#2{%
\if@filesw \immediate \write \@auxout {\string \citation {#2}}\fi 
\@tempcntb\m@ne \let\@h@ld\relax \def\@citea{}%
\@cite{%
  \@for \@citeb:=#2\do {%
    \@ifundefined {b@\@citeb}%
      {\@h@ld\@citea\@tempcntb\m@ne{\bf ?}%
      \@warning {Citation `\@citeb ' on page \thepage \space undefined}}%
      {\@tempcnta\@tempcntb \advance\@tempcnta\@ne%
      \@tempcntb\number\csname b@\@citeb \endcsname \relax%
      \ifnum\@tempcnta=\@tempcntb 
        \ifx\@h@ld\relax%
          \edef \@h@ld{\@citea\csname b@\@citeb\endcsname}%
        \else%
          \edef\@h@ld{\ifmmode{-}\else--\fi\csname b@\@citeb\endcsname}%
        \fi%
      \else
        \@h@ld\@citea\csname b@\@citeb \endcsname%
        \let\@h@ld\relax%
      \fi}%
    \def\@citea{,\penalty\@highpenalty\,}%
  }\@h@ld
}{#1}}
\def\@citey[#1]#2{%
\if@filesw \immediate \write \@auxout {\string \citation {#2}}\fi 
\@tempcntb\m@ne \let\@h@ld\relax \def\@citea{}%
\@icite{%
  \@for \@citeb:=#2\do {%
    \@ifundefined {b@\@citeb}%
      {\@h@ld\@citea\@tempcntb\m@ne{\bf ?}%
      \@warning {Citation `\@citeb ' on page \thepage \space undefined}}%
      {\@tempcnta\@tempcntb \advance\@tempcnta\@ne%
      \@tempcntb\number\csname b@\@citeb \endcsname \relax%
      \ifnum\@tempcnta=\@tempcntb 
        \ifx\@h@ld\relax%
          \edef \@h@ld{\@citea\csname b@\@citeb\endcsname}%
        \else%
          \edef\@h@ld{\ifmmode{-}\else--\fi\csname b@\@citeb\endcsname}%
        \fi%
      \else
        \@h@ld\@citea\csname b@\@citeb \endcsname%
        \let\@h@ld\relax%
      \fi}%
    \def\@citea{,\penalty\@highpenalty\,}%
  }\@h@ld
}{#1}}
\def\@cite#1#2{{$^{#1}$\if@tempswa , #2\fi }}
\def\@icite#1#2{{$#1$\if@tempswa , #2\fi }}

\gdef\@publabel{\hfil}
\gdef\@pubdate{\null}
\gdef\@pubnumber{\null}
\gdef\@author{\null}
\gdef\@title{\null}
\gdef\@abstract{\null}
\long\def\pubdate#1{\gdef\@pubdate{#1}}
\long\def\pubnumber#1{\gdef\@pubnumber{#1}}
\long\def\publabel#1{\gdef\@publabel{#1}}
\long\def\author#1{\gdef\@author{#1}}
\long\def\title#1{\gdef\@title{#1}}
\long\def\abstract#1{\gdef\@abstract{#1}}
\def\titlerelax{
\let\maketitle\relax
\let\settitleparameters\relax
\let\consolidatetitle\relax
\let\inittitlepage\relax
\let\finishtitlepage\relax
\let\titlepagecontents\relax
\let\multithanks\relax
\let\titlebaselines\relax
\let\@makepub\relax
\let\@maketitle\relax
\let\@makeauthor\relax
\let\@makeabstract\relax
\let\@maketitlenote\relax
\let\thanks\relax
\let\titlerelax\relax}

\def\titleclean
{\gdef\@titlenote{}
\gdef\@abstract{}
\gdef\@author{}
\gdef\@title{}
\gdef\@pubdate{}\gdef\@pubnumber{}\gdef\@publabel{}
\gdef\@dpublabel{}
}
\def\@makepub{\vbox to \z@{\hbox to \textwidth{\hfill
\@publabel \hfill
\llap{\parbox[t]{0.25\textwidth}{\raggedleft\@pubnumber}}}%
\vss}}
\def\@maketitle{\vskip 60pt \begin{center}
 {\LARGE \@title \par}
 \end{center}}
\def\@makeauthor{{%
\def\and{\smallskip {\normalsize \rm and\smallskip }}
\def\And{\medskip {\normalsize \rm and\\}\medskip}
\long\def\address##1{{\def\and{\\and\\}\medskip
				{\small \it \\##1\\}
}}
{\centering
 \vskip 3em
 \large \lineskip .75em
 \@author}
 \par}} 
\def\@makedate{\vskip 1.5em 
 {\raggedright \small \noindent\@pubdate \par}}
\def\@makeabstract{\vskip 1.5em
{\small 
\begin{center}
{\bf ABSTRACT\vspace{-.5em}\vspace{0pt}} 
\end{center}
\quotation \@abstract \endquotation}}
\def\maketitle{\titlepage
\let\footnotesize\small \setcounter{page}{0}
\def\thefootnote{\fnsymbol{footnote}}
\@makepub
\vfil
\@maketitle
\@makeauthor
\vfil
\@makeabstract
\@thanks
\vfil
\@makedate
\if@restonecol\twocolumn \else \eject \fi
\titlerelax \titleclean
\def\thefootnote{\alph{footnote}}
\setcounter{footnote}{0}
}

\ifcase\@ptsize
 \font\tenmsa=msam10
 \font\sevenmsa=msam7
 \font\fivemsa=msam5
 \font\tenmsb=msbm10
 \font\sevenmsb=msbm7
 \font\fivemsb=msbm5
\or
 \font\tenmsa=msam10 scaled \magstephalf
 \font\sevenmsa=msam8
 \font\fivemsa=msam6
 \font\tenmsb=msbm10 scaled \magstephalf
 \font\sevenmsb=msbm8
 \font\fivemsb=msbm6
\or
 \font\tenmsa=msam10 scaled \magstep1
 \font\sevenmsa=msam8
 \font\fivemsa=msam6
 \font\tenmsb=msbm10 scaled \magstep1
 \font\sevenmsb=msbm8
 \font\fivemsb=msbm6
\fi

\newfam\msafam
\newfam\msbfam
\textfont\msafam=\tenmsa  \scriptfont\msafam=\sevenmsa
  \scriptscriptfont\msafam=\fivemsa
\textfont\msbfam=\tenmsb  \scriptfont\msbfam=\sevenmsb
  \scriptscriptfont\msbfam=\fivemsb

\def\hexnumber@#1{\ifnum#1<10 \number#1\else
 \ifnum#1=10 A\else\ifnum#1=11 B\else\ifnum#1=12 C\else
 \ifnum#1=13 D\else\ifnum#1=14 E\else\ifnum#1=15 F\fi\fi\fi\fi\fi\fi\fi}

\def\msa@{\hexnumber@\msafam}
\def\msb@{\hexnumber@\msbfam}
\mathchardef\boxdot="2\msa@00
\mathchardef\boxplus="2\msa@01
\mathchardef\boxtimes="2\msa@02
\mathchardef\square="0\msa@03
\mathchardef\blacksquare="0\msa@04
\mathchardef\centerdot="2\msa@05
\mathchardef\lozenge="0\msa@06
\mathchardef\blacklozenge="0\msa@07
\mathchardef\circlearrowright="3\msa@08
\mathchardef\circlearrowleft="3\msa@09
\mathchardef\rightleftharpoons="3\msa@0A
\mathchardef\leftrightharpoons="3\msa@0B
\mathchardef\boxminus="2\msa@0C
\mathchardef\Vdash="3\msa@0D
\mathchardef\Vvdash="3\msa@0E
\mathchardef\vDash="3\msa@0F
\mathchardef\twoheadrightarrow="3\msa@10
\mathchardef\twoheadleftarrow="3\msa@11
\mathchardef\leftleftarrows="3\msa@12
\mathchardef\rightrightarrows="3\msa@13
\mathchardef\upuparrows="3\msa@14
\mathchardef\downdownarrows="3\msa@15
\mathchardef\upharpoonright="3\msa@16

\mathchardef\downharpoonright="3\msa@17
\mathchardef\upharpoonleft="3\msa@18
\mathchardef\downharpoonleft="3\msa@19
\mathchardef\rightarrowtail="3\msa@1A
\mathchardef\leftarrowtail="3\msa@1B
\mathchardef\leftrightarrows="3\msa@1C
\mathchardef\rightleftarrows="3\msa@1D
\mathchardef\Lsh="3\msa@1E
\mathchardef\Rsh="3\msa@1F
\mathchardef\rightsquigarrow="3\msa@20
\mathchardef\leftrightsquigarrow="3\msa@21
\mathchardef\looparrowleft="3\msa@22
\mathchardef\looparrowright="3\msa@23
\mathchardef\circeq="3\msa@24
\mathchardef\succsim="3\msa@25
\mathchardef\gtrsim="3\msa@26
\mathchardef\gtrapprox="3\msa@27
\mathchardef\multimap="3\msa@28
\mathchardef\therefore="3\msa@29
\mathchardef\because="3\msa@2A
\mathchardef\doteqdot="3\msa@2B

\mathchardef\triangleq="3\msa@2C
\mathchardef\precsim="3\msa@2D
\mathchardef\lesssim="3\msa@2E
\mathchardef\lessapprox="3\msa@2F
\mathchardef\eqslantless="3\msa@30
\mathchardef\eqslantgtr="3\msa@31
\mathchardef\curlyeqprec="3\msa@32
\mathchardef\curlyeqsucc="3\msa@33
\mathchardef\preccurlyeq="3\msa@34
\mathchardef\leqq="3\msa@35
\mathchardef\leqslant="3\msa@36
\mathchardef\lessgtr="3\msa@37
\mathchardef\backprime="0\msa@38
\mathchardef\risingdotseq="3\msa@3A
\mathchardef\fallingdotseq="3\msa@3B
\mathchardef\succcurlyeq="3\msa@3C
\mathchardef\geqq="3\msa@3D
\mathchardef\geqslant="3\msa@3E
\mathchardef\gtrless="3\msa@3F
\mathchardef\sqsubset="3\msa@40
\mathchardef\sqsupset="3\msa@41
\mathchardef\vartriangleright="3\msa@42
\mathchardef\vartriangleleft="3\msa@43
\mathchardef\trianglerighteq="3\msa@44
\mathchardef\trianglelefteq="3\msa@45
\mathchardef\bigstar="0\msa@46
\mathchardef\between="3\msa@47
\mathchardef\blacktriangledown="0\msa@48
\mathchardef\blacktriangleright="3\msa@49
\mathchardef\blacktriangleleft="3\msa@4A
\mathchardef\vartriangle="3\msa@4D
\mathchardef\blacktriangle="0\msa@4E
\mathchardef\triangledown="0\msa@4F
\mathchardef\eqcirc="3\msa@50
\mathchardef\lesseqgtr="3\msa@51
\mathchardef\gtreqless="3\msa@52
\mathchardef\lesseqqgtr="3\msa@53
\mathchardef\gtreqqless="3\msa@54
\mathchardef\Rrightarrow="3\msa@56
\mathchardef\Lleftarrow="3\msa@57
\mathchardef\veebar="2\msa@59
\mathchardef\barwedge="2\msa@5A
\mathchardef\doublebarwedge="2\msa@5B
\mathchardef\angle="0\msa@5C
\mathchardef\measuredangle="0\msa@5D
\mathchardef\sphericalangle="0\msa@5E
\mathchardef\varpropto="3\msa@5F
\mathchardef\smallsmile="3\msa@60
\mathchardef\smallfrown="3\msa@61
\mathchardef\Subset="3\msa@62
\mathchardef\Supset="3\msa@63
\mathchardef\Cup="2\msa@64

\mathchardef\Cap="2\msa@65

\mathchardef\curlywedge="2\msa@66
\mathchardef\curlyvee="2\msa@67
\mathchardef\leftthreetimes="2\msa@68
\mathchardef\rightthreetimes="2\msa@69
\mathchardef\subseteqq="3\msa@6A
\mathchardef\supseteqq="3\msa@6B
\mathchardef\bumpeq="3\msa@6C
\mathchardef\Bumpeq="3\msa@6D
\mathchardef\lll="3\msa@6E

\mathchardef\ggg="3\msa@6F

\mathchardef\circledS="0\msa@73
\mathchardef\pitchfork="3\msa@74
\mathchardef\dotplus="2\msa@75
\mathchardef\backsim="3\msa@76
\mathchardef\backsimeq="3\msa@77
\mathchardef\complement="0\msa@7B
\mathchardef\intercal="2\msa@7C
\mathchardef\circledcirc="2\msa@7D
\mathchardef\circledast="2\msa@7E
\mathchardef\circleddash="2\msa@7F
\def\ulcorner{\delimiter"4\msa@70\msa@70 }
\def\urcorner{\delimiter"5\msa@71\msa@71 }
\def\llcorner{\delimiter"4\msa@78\msa@78 }
\def\lrcorner{\delimiter"5\msa@79\msa@79 }
\def\yen{\mathhexbox\msa@55 }
\def\checkmark{\mathhexbox\msa@58 }
\def\circledR{\mathhexbox\msa@72 }
\def\maltese{\mathhexbox\msa@7A }
\mathchardef\lvertneqq="3\msb@00
\mathchardef\gvertneqq="3\msb@01
\mathchardef\nleq="3\msb@02
\mathchardef\ngeq="3\msb@03
\mathchardef\nless="3\msb@04
\mathchardef\ngtr="3\msb@05
\mathchardef\nprec="3\msb@06
\mathchardef\nsucc="3\msb@07
\mathchardef\lneqq="3\msb@08
\mathchardef\gneqq="3\msb@09
\mathchardef\nleqslant="3\msb@0A
\mathchardef\ngeqslant="3\msb@0B
\mathchardef\lneq="3\msb@0C
\mathchardef\gneq="3\msb@0D
\mathchardef\npreceq="3\msb@0E
\mathchardef\nsucceq="3\msb@0F
\mathchardef\precnsim="3\msb@10
\mathchardef\succnsim="3\msb@11
\mathchardef\lnsim="3\msb@12
\mathchardef\gnsim="3\msb@13
\mathchardef\nleqq="3\msb@14
\mathchardef\ngeqq="3\msb@15
\mathchardef\precneqq="3\msb@16
\mathchardef\succneqq="3\msb@17
\mathchardef\precnapprox="3\msb@18
\mathchardef\succnapprox="3\msb@19
\mathchardef\lnapprox="3\msb@1A
\mathchardef\gnapprox="3\msb@1B
\mathchardef\nsim="3\msb@1C
\mathchardef\napprox="3\msb@1D
\mathchardef\varsubsetneq="3\msb@20
\mathchardef\varsupsetneq="3\msb@21
\mathchardef\nsubseteqq="3\msb@22
\mathchardef\nsupseteqq="3\msb@23
\mathchardef\subsetneqq="3\msb@24
\mathchardef\supsetneqq="3\msb@25
\mathchardef\varsubsetneqq="3\msb@26
\mathchardef\varsupsetneqq="3\msb@27
\mathchardef\subsetneq="3\msb@28
\mathchardef\supsetneq="3\msb@29
\mathchardef\nsubseteq="3\msb@2A
\mathchardef\nsupseteq="3\msb@2B
\mathchardef\nparallel="3\msb@2C
\mathchardef\nmid="3\msb@2D
\mathchardef\nshortmid="3\msb@2E
\mathchardef\nshortparallel="3\msb@2F
\mathchardef\nvdash="3\msb@30
\mathchardef\nVdash="3\msb@31
\mathchardef\nvDash="3\msb@32
\mathchardef\nVDash="3\msb@33
\mathchardef\ntrianglerighteq="3\msb@34
\mathchardef\ntrianglelefteq="3\msb@35
\mathchardef\ntriangleleft="3\msb@36
\mathchardef\ntriangleright="3\msb@37
\mathchardef\nleftarrow="3\msb@38
\mathchardef\nrightarrow="3\msb@39
\mathchardef\nLeftarrow="3\msb@3A
\mathchardef\nRightarrow="3\msb@3B
\mathchardef\nLeftrightarrow="3\msb@3C
\mathchardef\nleftrightarrow="3\msb@3D
\mathchardef\divideontimes="2\msb@3E
\mathchardef\varnothing="0\msb@3F
\mathchardef\nexists="0\msb@40
\mathchardef\mho="0\msb@66
\mathchardef\thorn="0\msb@67
\mathchardef\beth="0\msb@69
\mathchardef\gimel="0\msb@6A
\mathchardef\daleth="0\msb@6B
\mathchardef\lessdot="3\msb@6C
\mathchardef\gtrdot="3\msb@6D
\mathchardef\ltimes="2\msb@6E
\mathchardef\rtimes="2\msb@6F
\mathchardef\shortmid="3\msb@70
\mathchardef\shortparallel="3\msb@71
\mathchardef\smallsetminus="2\msb@72
\mathchardef\thicksim="3\msb@73
\mathchardef\thickapprox="3\msb@74
\mathchardef\approxeq="3\msb@75
\mathchardef\succapprox="3\msb@76
\mathchardef\precapprox="3\msb@77
\mathchardef\curvearrowleft="3\msb@78
\mathchardef\curvearrowright="3\msb@79
\mathchardef\digamma="0\msb@7A
\mathchardef\varkappa="0\msb@7B
\mathchardef\hslash="0\msb@7D
\mathchardef\hbar="0\msb@7E
\mathchardef\backepsilon="3\msb@7F
\def\Bbb{\ifmmode\let\next\Bbb@\else
 \def\next{\errmessage{Use \string\Bbb\space only in math mode}}\fi\next}
\def\Bbb@#1{{\Bbb@@{#1}}}
\def\Bbb@@#1{\fam\msbfam#1}

\def\bk {{\hskip 0.2 cm}}

\def\com{{\hskip 0.2 cm},}
\def\pkt{{\hskip 0.2 cm}.}
\def\acknowledgements{\@startsection{section}{4}
{\z@}{-3.5ex plus -1ex minus -.2ex}{2.3ex plus .2ex}{\normalsize\bf}
{Acknowledgements}}

\def\sct {{\rm{\tt{sc}}(2)}}      
\def\tilt {\tilde{t}}		  
\newcommand{\ch}[1]{\makebox(0,0)[b]{\scriptsize$#1$}}  
\newcommand{\chtl}[1]{\makebox(0,0)[l]{\tiny$#1$}}  
\newcommand{\chtr}[1]{\makebox(0,0)[r]{\tiny$#1$}}  
\newcommand{\secn}[1]{{\sc Sec.}\,{\sf #1}}	  
\newcommand{\eq}[1]{{\sc Eq.}\,{\sf (#1)}}	  
\newcommand{\eqs}[1]{{\sc Eqs.}\,{\sf (#1)}}	  
\newcommand{\eqoth}[1]{{\sf (#1)}}  	  	  

\def\bbbz {\Bbb{Z}}		  
\def\bbbzh{\Bbb{Z}_{\frac{1}{2}}} 

\def\bbbn {\Bbb{N}}   	          
\def\bbbnh{\Bbb{N}_{\frac{1}{2}}} 
\def\bbbc {\Bbb{C}}

\newtheorem{definition}{Definition}[section]
\newtheorem{theorem}[definition]{Theorem}
\newtheorem{lemma}[definition]{Lemma}
\newtheorem{proposition}[definition]{Proposition}
\newcounter{defs}[section]

\newcommand{\be}{\begin{equation}}
\newcommand{\ee}{\end{equation}}
\newcommand{\bea}{\begin{eqnarray}}
\newcommand{\eea}{\end{eqnarray}}

\newcommand{\bdf}{\stepcounter{defs}\begin{definition}}
\newcommand{\edf}{\end{definition}}
\newcommand{\bth}{\stepcounter{defs}\begin{theorem}}
\newcommand{\eth}{\end{theorem}}
\newcommand{\blm}{\stepcounter{defs}\begin{lemma}}
\newcommand{\elm}{\end{lemma}}
\newcommand{\bpr}{\stepcounter{defs}\begin{proposition}}
\newcommand{\epr}{\end{proposition}}
\newcommand{\bprf}{Proof: }
\newcommand{\eprf}{\hfill $\blacksquare$ \\}
\newcounter{pics}
\newcommand{\bpic}[4]{\begin{center}\begin{picture}(#1,#2)(#3,#4)
\refstepcounter{pics}}
\renewcommand{\thepics}{{\sf\roman{pics}}}
\newcommand{\epic}[1]{\end{picture}\\
{\small {\sc Fig.} \thepics \bk #1} \end{center}}
\newcommand{\epicspl}{\end{picture}\\		
\addtocounter{pics}{-1}\end{center}}		

\renewcommand{\thefootnote}{\rm{\alph{footnote}}}
\newcounter{tabs}
\newcommand{\btab}[1]{\refstepcounter{tabs}\begin{center}
\begin{tabular}{#1}}
\renewcommand{\thetabs}{{\sf\alph{tabs}}}
\newcommand{\etab}[1]{\end{tabular}\\[1.5ex]
{\small {\sc Tab.} \thetabs \bk #1} \end{center}}

\def\noi {\noindent}
\newcommand{\nn}{\nonumber}
\newcommand{\cl}{\tt}		  

\newcommand{\ket}[1]{\left| {#1} \right\rangle}	
\newcommand{\spn}[1]{{\tt span}\{{#1}\}}	
\newcommand{\starprod}[2]{{\displaystyle 	
{\prod_{#1}^{#2} \!\!\! {}^{>} \;}}}		
\newcommand{\sgn}[1]{{\tt sgn}\{{#1}\}}		
\newcommand{\trmod}[2]{{\tt tr}_{#1}
\left( #2 \right) }					
\def\det{{\tt det}}				
\def\ker{{\tt kernel}}				

\def\lch{{\large $\lhd$}}			
\def\rch{{\large $\rhd$}}		     	
\newcommand{\ty}[1]{{\scriptsize #1}}           

\def\pmb#1{\setbox0=\hbox{#1}%
 \kern-.025em\copy0\kern-\wd0
 \kern.05em\copy0\kern-\wd0
 \kern-.025em\raise.0433em\box0 }

\makeatother

\title{The embedding structure of unitary\\
 $N=2$ minimal models}
\author{Matthias D\"{o}rrzapf
\thanks{e-mail: matthias@feynman.harvard.edu} 
\address{Lyman Laboratory of Physics\\
Harvard University\\
Cambridge, MA 02138, USA }}\pubdate{December 1997}
\pubnumber{HUTP-97/A056 \\ hep-th/9712165}

\abstract{
We derive the embedding structure of unitary $N=2$ minimal models 
and show as a result
that these representations
have a degeneration of uncharged singular states.
This corrects
some earlier mistakes made in the literature. 
We discuss the connexion
to the $N=2$ character formulae and 
finally give a proof for the embedding diagrams.}

\begin{document}

\maketitle



\section{Introduction}

Conformal and superconformal symmetry is the underlying structure of
many models in very different areas of physics. This includes
statistical systems, random walk models, percolation models
and also string theory. In particular, the $N=2$ superconformal 
algebras play a crucial r\^ole for string theory, since they supply
the underlying symmetry for the $N=2$ string.
The representation theory of the {\it Virasoro algebra}, the algebra of
generators of local two-dimensional conformal symmetry, 
is well-understood. Based on it, the generalisation to the $N=1$
superconformal algebras is straightforward. At first,
it seemed as if the representation theory of
$N=2$ superconformal algebras followed similar patterns and
character formulae as well as embedding diagrams were therefore
derived\cite{dobrev,kiritsis2,matsuo} applying methods 
used for the Virasoro algebra 
assuming they would work for the representations of the $N=2$
superconformal algebras as well. However, 
we have shown in Ref. \icite{paper2} and
Ref. \icite{thesis} that the $N=2$ algebras are more complicated than 
originally believed. We have proven that the existing 
embedding diagrams are wrong for some crucial models and we could show
that in some $N=2$ cases one obtains degenerated singular vectors.
The {\it discrete series} of unitary cases or {\it unitary minimal
models} is one class of representations
for which the previous embedding patterns fail. 

Starting from the $N=2$ determinant formula and taking 
the degenerated singular vector spaces into account, we
obtained a full set of embedding diagrams containing all the singular
vectors {\it guaranteed by the determinant formula}.
Our classification follows the {\it Feigin and Fuchs ``I-II-III pattern''}
and can be found in Ref. \icite{thesis}.
But the $N=2$ superconformal algebra turned out to be all the more
interesting as subsingular vectors were recently discovered
by Gato-Rivera and Rosado. Therefore it is not sufficient
to follow only the vanishing curves of the determinant 
formula in order to derive
the embedding structure. The unitary embedding diagrams, however,  
as we shall see at the end of 
this paper, do not contain 
any subsingular vectors and our results of Ref. \icite{thesis} hold
for at least these cases.
There are intriguing similarities between these $N=2$ embedding diagrams
and embedding diagrams of class IV of the affine Lie
superalgebra $\widehat{sl}(2,1;\bbbc)$, as shown in Ref. \icite{petanne}.
Just recently, the embedding diagrams of the $N=2$ superconformal
algebra were derived in an alternative way in Ref. \icite{semsir}. 
The authors of Ref. \icite{semsir} show that the
embedding structure of $\widehat{sl}(2)$ entails the embedding structure 
of the $N=2$ superconformal algebra and ultimately rederive many of our
results of Ref. \icite{thesis}.  

In this paper we present the embedding structure of the most important
highest weight representations: the unitary representations.
There are two classes of unitary representations, the 
{\it continuous class} and the {\it discrete series}.
The embedding diagrams for the continuous class given earlier by
Dobrev are correct\cite{dobrev}. The problems arise for the discrete series.
These are the cases we want to concentrate on.
After a short
introduction to the notation and some necessary conventions in 
\secn{2}, we review in \secn{3} some results of our Ref. \icite{paper2}
which are necessary for the reader to understand the difference
in structure between the Virasoro algebra and the $N=2$ algebra.
In \secn{4} we present the embedding structure of the
unitary minimal models which
is generalised to a larger class of  
representations in \secn{5}. In \secn{6}
we follow the idea of Eholzer and Gaberdiel\cite{wolfgmatth} to show
that the embedding diagrams imply the correct character formulae that
finally proves the embedding
diagrams of \secn{4} and \secn{5}.


\section{$N=2$ superconformal algebra and its unitary representations}

In a quantum field theory, unitarity is fundamental in that it is the condition
of conservation of probability. Hence, from the viewpoint
of a quantum field theory
the most interesting representations of the $N=2$ superconformal
algebras are the unitary
representations.
The unitary representations for the $N=2$ superconformal algebras
have been identified by
Boucher, Friedan and Kent\cite{adrian} for all different sectors of
the $N=2$ superconformal algebra. Similarly to the Virasoro
case, one finds a discrete series, called ``the unitary minimal models'',
and a continuous
class of unitary representations. 
The continuous class of unitary
highest weight representations has been analysed correctly by 
Dobrev\cite{dobrev}. Therefore we shall concentrate on the
discrete series of unitary representations.
Related to this, Eholzer and Gaberdiel demonstrated recently another
important feature of the $N=2$ algebra.
They showed in Ref. \icite{wolfgmatth} that all
{\it rational} $N=2$ superconformal theories
may be unitary. Where a theory is called rational
if it has only finitely many irreducible highest weight 
representations, and if the highest weight space of each of them
is finite dimensional. In \secn{6} we shall use their method of
deriving character formulae 
out of the embedding diagrams for the unitary minimal
models.

We denote the $N=2$ superconformal algebra in the
Neveu-Schwarz (or antiperiodic) moding by $\sct$.
It is given by the Virasoro algebra, the Heisenberg
algebra plus two anticommuting subalgebras with the
(anti-)commutation relations
\footnote{We write 
$\bbbn$ for $\{1,2,3,\ldots \}$, $\bbbn_{0}$
for $\{0,1,2,\ldots \}$, $\bbbn_{\frac{1}{2}}$ for 
$\{\frac{1}{2},\frac{3}{2},\frac{5}{2},\ldots \}$
and 
$\bbbz_{\frac{1}{2}}$ for 
$\{\ldots, -\frac{1}{2}, \frac{1}{2},\frac{3}{2},\ldots \}$.}:
\bea
[L_{m},L_{n}] & = & (m-n) L_{m+n} + \frac{C}{12} \:(m^{3}-m)\:
\delta_{m+n,0} \;\; ,\nn \\
\ [L_{m},G_{r}^{\pm}] & = & (\frac{1}{2} m-r) G_{m+r}^{\pm} \;\; ,\nn \\
\ [L_{m},T_{n}] & = & -n T_{m+n} \;\; ,\nn \\
\ [T_{m},T_{n}] & = & \frac{1}{3} C m \delta_{m+n,0}  
\;\; ,\label{eq:cr} \\
\ [T_{m},G_{r}^{\pm}] & = & \pm G_{m+r}^{\pm} \;\; ,\nn \\
\ \{ G_{r}^{+},G_{s}^{-}\} & = & 2 L_{r+s}+(r-s) T_{r+s} +\frac{C}{3}
(r^{2}-\frac{1}{4}) \delta_{r+s,0} \;\; ,\nn \\
\ [L_{m},C] & = & [T_{m},C] \;\; = \;\; [G_{r}^{\pm},C] 
\;\;=\;\; 0 \;\; ,\nn \\
\ \{G_{r}^{+},G_{s}^{+}\} & = & \{G_{r}^{-},G_{s}^{-}\}=0 , \;\;\;\;
\;\;\;\; m,n \in \bbbz , \;\; r,s \in \bbbz_{\frac{1}{2}} \;\; .\nn 
\eea
We can write $\sct$ in its triangular decomposition:
$\sct=\sct_{-} 
\oplus {\cal H}_{2} \oplus \sct_{+}$, where 
$\sct_{\pm}=\spn{ L_{\pm n},T_{\pm n},G_{\pm r}^{+},G_{\pm r}^{-}: n
\in \bbbn, r \in \bbbn_{\frac{1}{2}} }$, and
${\cal H}_{2}=\spn{L_{0},T_{0},C}$ is the
{\it grading preserving} Cartan subalgebra\footnote{There are four-dimensional
abelian subalgebras of $\sct$. However, these triangular decompositions
of the algebra are not consistent with our $\bbbz_2$-grading.}.
A simultaneous eigenvector
$\ket{h,q,c}$ of ${\cal H}_{2}$ with $L_{0}$, $T_{0}$ and $C$
eigenvalues $h$,
$q$ and $c$ respectively
and vanishing $\sct_{+}$ action $\sct_{+}
 \ket{h,q,c} =0$,
is called a highest weight vector. The Verma module ${\cal V}_{h,q,c}$
is defined as the $\sct$ left module $U(\sct)
 \otimes_{{\cal
H}_{2} \oplus \sct_{+}} \ket{h,q,c}$,
 where $U(\sct)$ denotes the
universal enveloping algebra of $\sct$. 
Finally, we call a vector
singular in ${\cal V}_{h,q,c}$, if it is not proportional to the highest
weight vector but still satisfies the highest weight vector 
conditions:
 $\Psi_{n,p} \in
{\cal V}_{h,q,c}$ is called singular if $L_{0}
\Psi_{n,p} =(h+n) \Psi_{n,p}$, 
$T_{0}
\Psi_{n,p} =(q+p) \Psi_{n,p}$ and
$\sct_{+} \Psi_{n,p} =0$ for some $n \in \bbbn$ and $p \in \bbbz$.
 If a vector is an
eigenvector of $L_{0}$ we call its eigenvalue $h$ its {\sl conformal
weight} and similarly its eigenvalue of $T_{0}$ is called its 
{\sl $U(1)$-charge}\footnote{For a singular vector 
$\Psi_{n,p}\in {\cal V}_{h,q,c}$ 
we may simply say its {\it charge} $p$ and its {\it level} $n$ rather than 
$U(1)$-charge $q+p$ and conformal weight $h+n$.}. 

The determinant formula given by Boucher, Friedan and Kent\cite{adrian} 
makes it apparent that the Verma module ${\cal V}_{h_{r,s}(t,q),q,c(t)}$ has
for positive, integral $r$ and positive,
even $s$ an uncharged singular vector 
at level
$\frac{rs}{2}$ which we shall call $\Psi_{r,s}=\Theta_{r,s}\ket{h_{r,s},
q,c}$, with $\Theta_{r,s}\in\sct_{-}$.
We use the parametrisation
\bea
c(t) = 3-3t  &\;\; ,& 
h_{r,s}(t,q) = \frac{(s-rt)^{2}}{8t} - \frac{q^{2}}{2t} -
\frac{t}{8} \;\; . \label{eq:param1}
\eea
We can find $\pm 1$ 
charged singular vectors $\Psi_{k}^{\pm}=\Theta^{\pm}_k \ket{h^{\pm}_{k},
q,c}$ (with $\Theta^{\pm}_k \in \sct_{-}$) in the Verma module
${\cal V}_{h_{k}^{\pm}(t,q),q,c(t)}$ at level $k$ for $k \in
\bbbn_{\frac{1}{2}}$.
The conformal weight $h_{k}^{\pm}$ is
\bea
h_{k}^{\pm}(t,q) &=& \pm k q +\frac{1}{2} t (k^{2}-\frac{1}{4}) \; .
\label{eq:param2}
\eea
For convenience we shall frequently use $\tilde t=\frac{t}{2}$ and
$\tilde s=\frac{s}{2}$. If we use the parametrisation 
\bea
a^2=4\tilde t h +\tilde t^2 +q^2 \com & 
{\displaystyle k^a=-\frac{q}{2\tilde t}+\frac{a}{2\tilde t}} \com & 
k^b=-\frac{q}{2\tilde t}-\frac{a}{2\tilde t} \com
\eea
then the determinant expression factorises (for $t\neq 0$) 
proportional to the following product ($n$ refers to the level and
$m$ to the charge):
\bea
\det{M_{n,m}} \propto \underbrace{\prod_{1\leq r\tilde{s}\leq n \atop
r,\tilde{s} \in \bbbz} (\tilde{s}-r\tilde{t}+a)^{P(n-r\tilde{s},m)}}_{
\rm{uncharged} \; \rm{sector}}
\underbrace{\prod_{k\in \bbbzh} 
\left[(k-k^{a})(k-k^{b})\right]^{\tilde{P}
(n-\mid k \mid,m-\sgn{k},k)}}_{\rm{charged} \; \rm{sector}} \; ,
\label{eq:det}
\eea
where $P$ and $\tilde P$ are the corresponding 
{\it partition functions}\cite{adrian}.
In order to analyse the singular vectors {\it guaranteed by the
determinant}, we simply need to find integer solutions $(r,\tilde s)$ on
the straight line $\tilde s=r\tilde t-a$ for the uncharged sector
leading us to $\Psi_{r,s}$ and
we need to verify if $k^a$ or $k^b$ is half-integral for the charged 
sector. If $k^a$ or $k^b$ is $\in\bbbz_{\frac{1}{2}}$,
then there is a $\sgn{k^a}$ or $\sgn{k^b}$
charged singular vector at level $|k^a|$ or $|k^b|$
respectively. Once we have found singular vectors in a Verma
module, we can then analyse the submodules built on top
of these singular vectors by computing the new parameters $a$, $k^a$ and
$k^b$ for the embedded modules. However, singular vectors of 
embedded modules may be trivial in the original Verma module. Using 
\eqs{\ref{eq:van1}-\ref{eq:van3}} of the following section, 
allows us to verify whenever this happens.

As shown by Gato-Rivera and Rosado\cite{beatriz1}, representations
of the $N=2$ 
superconformal algebra
contain in some cases vectors 
which become singular in the quotient module of 
the Verma module with a submodule but are not singular in the original
Verma module. This kind of vectors is called {\it subsingular
vectors}. Each submodule of a Verma module is generated by singular and 
subsingular vectors. 

In the {\it Neveu-Schwarz sector}
of the algebra, for an element of the discrete series 
of unitary representations there exists a
number $m \in
\bbbn, m\geq 2$ such that $t=\frac{2}{m}$. Besides, there exist
two half-integral numbers $j,k \in \bbbnh$ and $0<j,k,j+k\leq m-1$ such
that the conformal weight is given by $h=\frac{jk-\frac{1}{4}}{m}$ and
the $U(1)$-charge by $q=\frac{j-k}{m}$.

Schwimmer and Seiberg showed\cite{schwimmer} that the
$N=2$ Ramond algebra is simply a rewriting of the $N=2$ Neveu-Schwarz
algebra using the algebra isomorphism ${\cal U}_{\theta}$ called the
{\it spectral flow}.
\bea
{\cal U}_{\theta} L_{m} {\cal U}_{-\theta} &=& L_m + \theta T_m
+ \frac{c}{6} \theta^2 \delta_{m,0} \nn \\
{\cal U}_{\theta} T_{m} {\cal U}_{-\theta} &=& T_m 
+ \frac{c}{3} \theta \delta_{m,0} \\
{\cal U}_{\theta} G^\pm_{m} {\cal U}_{-\theta} &=& G^\pm_{m\pm \theta} \nn
\eea 
The Neveu-Schwarz and the Ramond algebras are connected
via the spectral flow using $\theta=\frac{1}{2}$ or 
$\theta=-\frac{1}{2}$.

Ramond highest weight representations consist of two independent 
sectors (the $+$ and the $-$ sector). 
Each of them built on highest wight vectors satisfying
an additional condition that $G^+_0$ or $G^-_0$ annihilates the
highest weight vector. Only for $h = \frac{c}{24}$ we find 
highest weight vectors satisfying both constraints but also 
highest weight vectors satisfying none of the additional
constraints. The latter ones 
do not need to be considered for unitary representations
since they contain level $0$ singular states. Dividing these out
leads us again to the $+$ and $-$ sectors. 

The discrete series of 
unitary representations in the Ramond $\pm$ sectors are given\cite{adrian} by
integers $m\geq 2$, $J$ and $K$ such that $t=\frac{2}{m}$, 
$h = \frac{JK}{m} + \frac{c}{24}$ and $q=\pm \frac{J-K}{m}$
with $0\leq J-1, K, J+K\leq m-1$. It can easily be verified that
the spectral flow
maps the unitary representations of the Ramond and of the Neveu-Schwarz
algebra onto each other such that $J=j+\frac{1}{2}$ and $K=k-\frac{1}{2}$.
The Ramond embedding diagrams coincide with the ones for the Neveu-Schwarz
algebra, taking into account that the level of the charged
singular vectors are shifted by $\frac{1}{2}$ and therefore
some of the uncharged singular vectors may be at the same level
as the charged singular vectors although in the case of the
Neveu-Schwarz algebra
they appear at different levels. 
How to use the topological twists in order to transform
singular vector embedding structures from the Neveu-Schwarz
sector to the topological algebra is described in Ref.
\icite{beatriz1}. We shall therefore focus on the 
Neveu-Schwarz sector.


\section{Fermionic uncharged singular vectors}

In Ref. \icite{paper2} we showed that the Verma modules ${\cal V}_{h,q,t}$
are embedded in the analytically continued space $\widetilde{\cal V}_{h,q,t}$
in which the operators $L_{-1}$ are continued to arbitrary complex
powers $L_{-1}^{a}$, for $a\in \bbbc$. $L_{-1}^{a}$ together with
$\sct_-$ generates $\widetilde{\sct_-}$. We therefore obtain
the generalised algebra $\widetilde{\sct} = \widetilde{\sct}_{-}
\oplus {\cal H}_{2} \oplus \sct_{+}$. The generalised 
(anti-)commutation
relations\cite{paper2} of $\widetilde{\sct}$ contain infinite but countable
sums in powers of $L_{-1}$.
The extended Verma modules 
$\widetilde{{\cal V}}_{h,q,c}= U(\widetilde{\sct}) 
\otimes_{{\cal H}_{2} \oplus \sct_{+}} \ket{h,q,c}$ 
decomposes just like the original Verma module ${\cal V}_{h,q,c}$ in 
infinitely many $(L_0,T_0)$-grade spaces, 
however, this time the grade spaces themselves
are infinite dimensional and furthermore the level of a grade
space is given by a complex number and there is a 
non-trivial infinite dimensional grade space
for each complex number.

A generalised singular vector\cite{paper2} 
in $\widetilde{{\cal V}}_{h,q,c}$ is a vector
such that its cut off vectors of order $M$ satisfy
the highest weight conditions up to order $M$, for all $M\in \bbbn$.
A singular vector $\Psi\in  {\cal V}_{h,q,t}$ is therefore also singular
in this generalised space and conversely the finite generalised
singular vectors $\tilde\Psi^f\in \widetilde{{\cal V}}_{h,q,c}$
are exactly the ones satisfying $\sct_+ \tilde\Psi^f \equiv 0$.
However, this does not necessarily mean that $\tilde\Psi^f \in
{\cal V}_{h,q,c}$, but if it was the case that 
$\tilde\Psi^f \in{\cal V}_{h,q,c}$
then $\tilde\Psi^f$ would be singular
in ${\cal V}_{h,q,c}$.

For $t\neq 0$ the singular
vectors of $\widetilde{{\cal V}}_{h,q,c}$ are all known. They
can be constructed by using products of analytic continuations\cite{paper2} of
the operators $\Theta_k^{\pm}$ for values of $k\in \bbbc$. 
We find that the space of uncharged
singular vectors in $\widetilde{{\cal V}}_{h,q,c}$ is exactly 
two-dimensional for each level $a\in \bbbc$. This two-dimensional space is 
spanned by\footnote{Since the order in the product is
significant, we define $\starprod{m=a}{b} f(m) = f(b) f(b-1) \ldots
f(a+1) f(a)$. } 
\bea
\Delta_{r,s}(0,1) \! &=&\! \frac{1}{2^{\frac{s}{2}-1}}
 \starprod{m=0}{\frac{s}{2}-1}
\Theta^{-}_{-\frac{s-rt}{2t}+\frac{q}{t}+\frac{2+2m}{t}}(t,q+1)
\Theta^{+}_{-\frac{s-rt}{2t}-\frac{q}{t}+\frac{2m}{t}}(t,q)
\ket{h_{r,s},q,c} \; , \label{eq:psi_rs1} \\
\Delta_{r,s}(1,0) \! &=&\! \frac{1}{2^{\frac{s}{2}-1}}
\starprod{m=0}{\frac{s}{2}-1}
\Theta^{+}_{-\frac{s-rt}{2t}-\frac{q}{t}+\frac{2+2m}{t}}(t,q-1)
\Theta^{-}_{-\frac{s-rt}{2t}+\frac{q}{t}+\frac{2m}{t}}(t,q)
\ket{h_{r,s},q,c} \; , \label{eq:psi_rs2}
\eea
for any complex numbers
$r,s$ such that $\frac{rs}{2}=a$. For the charged singular
vectors we find at each level in $\bbbc$
that there is an exactly one-dimensional vector space of
+1 charged singular vectors and a one-dimensional vector space 
of -1 charged singular vectors. These vectors can 
easily be constructed by using products of the generalised
operators $\Theta^{\pm}_k$ together with the vectors $\Delta_{r,s}(0,1)$ and 
$\Delta_{r,s}(1,0)$. This implies that in
${\cal V}_{h,q,c}$ there are at most two linearly independent uncharged
singular vectors at the same level and also $\pm 1$ charged
singular vectors are unique modulo scalar multiples. 
There cannot be any singular vectors 
in ${\cal V}_{h,q,c}$ with charges different to $0$ and $\pm 1$.

Using the leading terms of the singular vectors $\Psi_{r,s}\in
{\cal V}_{h_{r,s},q,c}$ we can identify these vectors among the
generalised singular vectors ($r\in\bbbn , s\in 2\bbbn$):
\bea
\Psi_{r,s}
&=& \epsilon^{+}_{r,s}(t,q) \theta_{r,s}(1,0) \ket{h_{r,s},q,c} +
\epsilon^{-}_{r,s}(t,q) \theta_{r,s}(0,1) \ket{h_{r,s},q,c}
\;\; , \label{eq:psieps}
\eea
with 
\bea
\epsilon^{\pm}_{r,s}(t,q) &=&\prod_{n=1}^{r} 
( \pm \frac{s-rt}{2t}+\frac{q}{t}\mp 
\frac{1}{2} \pm n)  \;\; , \label{eq:eps}
\eea
\noi and $\theta_{r,s}(1,0), \theta_{r,s}(0,1) 
\in \widetilde{\sct_{-}}$ defined in such a way
that $\theta_{r,s}(1,0)\ket{h_{r,s},q,c}=\Delta_{r,s}(1,0)$ and
 $\theta_{r,s}(0,1)\ket{h_{r,s},q,c}=\Delta_{r,s}(0,1)$.

$\Delta_{r,s}(1,0)$ and $\Delta_{r,s}(0,1)$ are in general not finite
and therefore they are 
certainly not singular vectors in ${\cal V}_{h,q,c}$. However,
their sum in \eq{\ref{eq:psieps}} has to be finite and in ${\cal V}_{h,q,c}$.
As a consequence, whenever $\epsilon^{+}_{r,s}(t,q)$ is trivial
we find that $\Delta_{r,s}(0,1) \in {\cal V}_{h,q,c}$
and thus $\Delta_{r,s}(0,1)$ is 
singular in ${\cal V}_{h,q,c}$.
Similarly, whenever $\epsilon^{-}_{r,s}(t,q)$ is 
trivial we have  $\Delta_{r,s}(1,0)$ singular in ${\cal V}_{h,q,c}$.
At the intersection points of $\epsilon^{+}_{r,s}(t,q)=0$ 
and\footnote{We could also imagine these conditions for different
pairs $(r,s)$ and $(\tilde r,\tilde s)$ if by coincidence 
$rs=\tilde r \tilde s$.}
$\epsilon^{-}_{r,s}(t,q)=0$, however, both 
$\Delta_{r,s}(1,0)$ and $\Delta_{r,s}(0,1)$ are singular in
${\cal V}_{h,q,c}$ and we obtain two singular vectors at the
same level and charge\cite{paper2}. In the following section, we
will show that most singular vectors of the discrete series of unitary 
representations
satisfy this condition. 
Therefore the embedding structure of the unitary minimal models
always shows this degeneracy of uncharged
singular vectors. The converse, that the unitary minimal models
are the only
ones showing this degeneracy is not true.

One of the major differences of the superconformal algebras
to the Virasoro algebra is that the product of two  
operators may be trivial due to the anticommuting property
of the fermionic operators.
For the understanding of the embedding
diagrams it is of particular interest to know whether or not products of
singular vector operators are trivial. In the Virasoro case
products of singular vector operators are never trivial and one
obtains the embedding diagrams simply by descending down from
low level singular vectors that are guaranteed by the determinant formula.

In Ref. \icite{paper2} we obtained multiplication rules
for singular vector operators. 
Using these rules and taking 
into account that $\theta(0,0)\equiv 0$ we find:
\bea
 \Theta^+_{k^+} \theta_{r^+,s^+} (1,0) \equiv 0 & \com &
\Theta^-_{k^-} \theta_{r^-,s^-} (0,1) \equiv 0 \com \label{eq:van1} \\
 \theta_{r^+,s^+} (0,1) \Theta^+_{j^+} \equiv 0 & \com &
\theta_{r^-,s^-} (1,0) \Theta^-_{j^-} \equiv 0 \com \label{eq:van2} \\
 \theta_{r_2,s_2} (1,0) \theta_{r_1,s_1} (0,1) = 0 & \com & 
\theta_{r_2,s_2} (0,1) \theta_{r_1,s_1} (1,0) = 0 \com \label{eq:van3}
\eea
assuming that the product of singular vector operators is defined: 
$h^\pm_{k^\pm}(t,q)=h_{r^\pm,s^\pm}(t,q)+\frac{r^\pm s^\pm}{2}$,
$h_{r^\pm,s^\pm}(t,q\pm 1)=h^\pm_{j^\pm}(t,q)+j^{\pm}$
or $h_{r_2,s_2}(t,q)=h_{r_1,s_1}(t,q)+\frac{r_1 s_1}{2}$ respectively.

Uncharged singular vectors of the type $\Delta_{r,s}(1,0)$ are thus 
annihilated by singular vector operators $\Theta^{+}_{k}$ provided the 
corresponding weight relation for $h_{r,s}$ and $h^+_{k}$ is satisfied. 
Conversely, the $+1$ charged descendant singular vector 
of a $\Psi^{-}_{j}$ is uncharged and of type
$\Delta(1,0)$. We shall call a vector of type $\Delta(1,0)$ 
{\it left fermionic} uncharged singular vector. Similarly,
the vectors of type $\Delta(0,1)$ are called
{\it right fermionic} uncharged singular vector.
The embedded Verma module built on a fermionic uncharged
singular vector is not complete. Therefore, it is of particular
interest to distinguish fermionic uncharged singular vectors from
``normal'' uncharged singular vectors in the embedding diagrams. 
We denote left or right fermionic uncharged 
singular vectors in embedding diagrams
and formulae
by $\lhd$ or $\rhd$ respectively.

Charged singular vectors are trivially fermionic
in the sense that they always satisfy condition \eqoth{\ref{eq:van2}}. 
In addition, we know that there are no charge $\pm2$ singular 
vectors. Therefore, for cases where the determinant formula
would naively imply a $\pm 2$ charged singular vector descending
from $\Psi^{\pm}_{k}$ by the operator $\Theta^{\pm}_{j}$, then we
know that $\Theta_{j}^{\pm}\Psi_{k}^{\pm}\equiv 0$.
It is a simple exercise to verify that if $\Psi^{\pm}_{k}$ is guaranteed by
the determinant formula (either directly or indirectly as descendant) with
parameter $k^{a}=\pm l$ or $k^b=\pm l$, 
then there exists an operator $\Theta^{\pm}_{l}$ which annihilates
$\Psi_{k}^{\pm}$. This is due to the fact that the module
built on top of $\Psi_{k}^{\pm}$ has again $k^{a}=\pm l$ or $k^b=\pm l$.
The set of operators annihilating $\Psi_{k}^{\pm}$ defines a 
submodule in ${\cal V}_{h^{\pm}_k+k,q\pm 1,t}$, which we shall
call the {\it kernel of}\footnote{Note that $\sct \Psi_{k}^{\pm}
=\frac{{\cal V}_{h^{\pm}_k+k,q\pm 1,t}}{\ker{\Psi_{k}^{\pm}}}$
is embedded in ${\cal V}_{h^{\pm}_k,q,t}$.}
$\Psi^{\pm}_k$.
As a submodule of ${\cal V}_{h^{\pm}_k+k,q\pm 1,t}$,
the kernel of $\Psi^{\pm}_k$ is generated by
singular and subsingular vectors of ${\cal V}_{h^{\pm}_k+k,q\pm 1,t}$.
It may even be generated by
a singular vector with relative level smaller than $l$.
In the unitary cases this does not happen. It will be crucial
for our later considerations to show that there are neither any
subsingular vectors in ${\cal V}_{h^{\pm}_k+k,q\pm 1,t}$ for the unitary 
minimal models.

A module generated by a charged singular vector is
smaller than a module generated by an uncharged singular vector 
without any constraints. The partition functions for such modules
can be found in Ref. \icite{adrian}. We shall need them in \secn{6}.


\section{Embedding diagrams}

Let us consider an element of the discrete series given by the parameters
$j$, $k$, and $m$, with $m\in\bbbn$,
$m\geq2$, such that $j,k \in\bbbnh$, $\tilt=\frac{1}{m}$ and
$0<j,k,j+k\leq m-1$,
then the conformal weight is given by 
$h=\frac{jk-\frac{1}{4}}{m}$ and the $U(1)$-charge by $q=
\frac{j-k}{m}$. 
We easily find for the parameter $a=\frac{j+k}{m}$ and 
$k^{a}=k\in\bbbnh$, $k^{b}=-j\in-\bbbnh$.
In the notation of Ref. \icite{thesis} these cases
are always
of type $\cl{III}_{-}^{0}A^{+}_{-}B^{-}_{+}$.

\bth \label{th:disrcser}
The unitary minimal models of the $\sct$ algebra
have the following embedding structure:
\eth
\vbox{
\bpic{200}{180}{0}{10} \label{pic:discrser}
\put(0,195){\ch{q:}}
\put(100,195){\ch{0}}
\put(50,195){\ch{-1}}
\put(150,195){\ch{+1}}
\put(98,180){\framebox(4,3){}}
\put(100,180){\line(0,-1){30}}
\put(100,150){\circle*{3}}
\put(102,152){\chtl{\Psi_{r_{-1},s_{-1}}}}
\put(153,152){\chtl{m\!-\!j\!-\!k}}
\put(100,150){\line(0,-1){30}}
\put(97,120){\circle*{3}}
\put(97,120){\makebox(0,0){\lch}}
\put(103,120){\circle*{3}}
\put(103,120){\makebox(0,0){\rch}}
\put(107,120){\chtl{\Psi_{r_1,s_1}}}
\put(153,120){\chtl{m\!+\!j\!+\!k}}
\put(97,120){\line(0,-1){30}}
\put(103,120){\line(0,-1){30}}
\put(97,90){\circle*{3}}
\put(103,90){\circle*{3}}
\put(97,90){\makebox(0,0){\lch}}
\put(103,90){\makebox(0,0){\rch}}
\put(107,90){\chtl{\Psi_{r_{2},s_{-2}}}}
\put(153,90){\chtl{2(2m\!-\!j\!-\!k)}}
\put(97,90){\line(0,-1){30}}
\put(103,90){\line(0,-1){30}}
\put(97,60){\circle*{3}}
\put(103,60){\circle*{3}}
\put(97,60){\makebox(0,0){\lch}}
\put(103,60){\makebox(0,0){\rch}}
\put(107,60){\chtl{\Psi_{r_2,s_2}}}
\put(153,60){\chtl{2(2m\!+\!j\!+\!k)}}
\put(97,60){\line(0,-1){30}}
\put(103,60){\line(0,-1){30}}
\put(97,30){\circle*{3}}
\put(103,30){\circle*{3}}
\put(97,30){\makebox(0,0){\lch}}
\put(103,30){\makebox(0,0){\rch}}
\put(107,30){\chtl{\Psi_{r_{-3},s_{-3}}}}
\put(153,30){\chtl{3(3m\!-\!j\!-\!k)}}
\put(50,160){\circle*{3}}
\put(47,162){\chtr{\Psi^-_j=\Psi^-_{r^-_0,s^-_0}; \; j}}
\put(150,160){\circle*{3}}
\put(153,162){\chtl{\Psi^+_k=\Psi^+_{r^+_0,s^+_0}; \; k}}
\put(50,159){\line(0,-1){28}}
\put(150,159){\line(0,-1){28}}
\put(50,130){\circle{3}}
\put(47,132){\chtr{\Psi^-_{r^-_{-2},s^-_{-2}}; \; 2m-j-2k}}
\put(150,130){\circle{3}}
\put(153,132){\chtl{\Psi^+_{r^+_{-2},s^+_{-2}}; \;2m-2j-k}}
\put(50,129){\line(0,-1){28}}
\put(150,129){\line(0,-1){28}}
\put(50,100){\circle{3}}
\put(47,102){\chtr{\Psi^-_{r^-_{1},s^-_{1}}; \; 2m+2j+k}}
\put(150,100){\circle{3}}
\put(153,102){\chtl{\Psi^+_{r^+_{1},s^+_{1}}; \; 2m+j+2k}}
\put(50,99){\line(0,-1){28}}
\put(150,99){\line(0,-1){28}}
\put(50,70){\circle{3}}
\put(47,72){\chtr{\Psi^-_{r^-_{-3},s^-_{-3}}; \; 6m-2j-3k}}
\put(150,70){\circle{3}}
\put(153,72){\chtl{\Psi^+_{r^+_{-3},s^+_{-3}}; \; 6m-3j-2k}}
\put(50,69){\line(0,-1){28}}
\put(150,69){\line(0,-1){28}}
\put(50,40){\circle{3}}
\put(47,42){\chtr{\Psi^-_{r^-_{2},s^-_{2}}; \; 6m+3j+2k}}
\put(150,40){\circle{3}}
\put(153,42){\chtl{\Psi^+_{r^+_{2},s^+_{2}}; \; 6m+2j+3k}}
\put(100,180){\line(5,-2){49}}
\put(100,180){\line(-5,-2){49}}
\put(100,150){\line(5,-2){49}}
\put(100,150){\line(-5,-2){49}}
\put(103,118){\line(5,-2){46}}
\put(97,118){\line(-5,-2){46}}
\put(103,88){\line(5,-2){46}}
\put(97,88){\line(-5,-2){46}}
\put(103,58){\line(5,-2){46}}
\put(97,58){\line(-5,-2){46}}
\put(103,122){\line(5,4){46}}
\put(97,122){\line(-5,4){46}}
\put(103,92){\line(5,4){46}}
\put(97,92){\line(-5,4){46}}
\put(103,62){\line(5,4){46}}
\put(97,62){\line(-5,4){46}}
\put(103,32){\line(5,4){46}}
\put(97,32){\line(-5,4){46}}
\put(97,20){\makebox(0,0){$\vdots$}}
\put(103,20){\makebox(0,0){$\vdots$}}
\put(50,30){\makebox(0,0){$\vdots$}}
\put(150,30){\makebox(0,0){$\vdots$}}
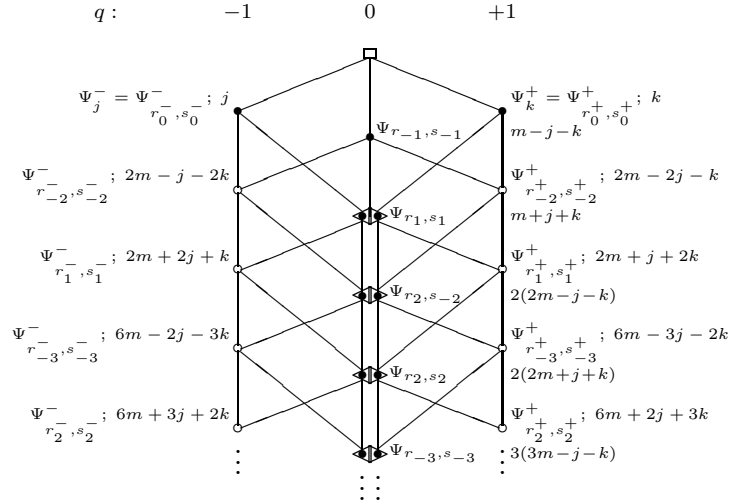
\epic{Discrete series of unitary representations: 
singular vectors and their levels.}}

We split the proof of theorem (\ref{th:disrcser}) into two parts.
We will spend most of this section to show that the embedding 
diagram necessarily contains the structure of Fig. \ref{pic:discrser}. 
That this structure
is also sufficient will be the result of \secn{6}.
In order to show necessity, we first analyse the singular vector
structure that is guaranteed by the determinant formula \eq{\ref{eq:det}}.
Factorising the determinant expression leads to uncharged 
singular vectors for integer pairs $(r,\tilde{s})$ on the
line $\tilde{s}=r\tilde{t}-a$ with $\tilde{t}=\frac{1}{m}$ and
$a=\frac{j+k}{m}$. The relevant integer points are 
thus $r_n=j+k+nm$, $\tilde{s}_n=n$ for $n\in\bbbz\setminus\{0\}$.
Therefore, we find uncharged singular vectors $\Psi_{r_n,s_n}$ at the levels
$r_n\tilde{s}_n$. As the given straight line intersects the
$r$-axis at an integer point, we find that descendant
singular vectors imply {\it secondary} singular vectors at exactly
the same levels. We shall demonstrate this for descendants of the singular
vector $\Psi_{r_{-1},s_{-1}}$ 
at level $r_{-1}\tilde{s}_{-1}=m-j-k$.
Factorising the determinant for $\Psi_{r_{-1},s_{-1}}$
as highest weight state gives
uncharged singular vectors for integer points on the straight line
$\tilde{s}^\prime=r^\prime\tilde{t}-a^\prime$ with $a^\prime=2-\frac{j+k}{m}$.
We then obtain descendant singular vectors of $\Psi_{r_{-1},s_{-1}}$
at the levels $m-j-k+r^\prime_n\tilde{s}^\prime_n$ with 
$r^\prime_n=2m-j-k+nm$ and $\tilde{s}^\prime_n=n$, where
$n\in\bbbz\setminus\{-2,-1,0\}$. Obviously, these levels agree with
$r_{n}\tilde{s}_n$. This should remind us of the Virasoro case\cite{ff1},
although here for the $N=2$ case 
the fact that the levels are the same does not
imply that these singular vectors are linearly
dependent, since the space of uncharged singular vectors can
be two-dimensional. We therefore need to analyse the dimensions
of the singular vector spaces. This is done by
verifying \eqs{\ref{eq:eps}}. If both 
$\epsilon^+_{r_n,\tilde{s}_n}(t,q)$ and $\epsilon^-_{r_n,\tilde{s}_n}(t,q)$
vanish,
then the space of the singular vectors guaranteed by the
determinant expression is two-dimensional. In addition, it would be
spanned by a left fermionic and a right fermionic uncharged singular vector.
Substituting the solutions $r_n$, $\tilde{s}_n$ into 
\eqs{\ref{eq:eps}} shows that both expressions vanish for $n\in\bbbn$,
but that they are non-trivial for $n\in - \bbbn$. We thus obtain
in the uncharged sector alternately a one-dimensional and a two-dimensional
singular vector space at levels $n(nm-j-k)$ and $n(nm+j+k)$
respectively, derived from the determinant formula for $n\in\bbbn$.
However, if we now perform the same calculation for the descendant
singular vectors of $\Psi_{r_{-1},s_{-1}}$, we find that the singular vector
spaces which were before one-dimensional become two-dimensional and vice 
versa. We note that $\Psi_{r_{-1},s_{-1}}$ is neither
left fermionic nor right fermionic. 
As a consequence,
all the singular vector spaces at the levels $r_n \tilde{s}_n$ are
two-dimensional in ${\cal V}_{h,q,c}$
and contain one left fermionic and one right
fermionic uncharged
singular vector with the single exception of $\Psi_{r_{-1},s_{-1}}$,
which stays one-dimensional and is not uncharged fermionic.

Let us now consider the charged singular vectors. $k^a=k$ and 
$k^b=-j$ show that there is a $-1$ charged singular vector at
level $j$ and a $+1$ charged singular vector at level $k$.
Descendant singular vectors of $\Psi^+_k$ in the $+1$ charged
sector correspond to integer
points on the straight line $\tilde{s}^+=r^+\tilde{t}-a^+$, where
$a^+=a+1$. The solutions to this problem satisfy again similar
fermionic conditions $\epsilon^{\pm}_{r^+,s^+}$. However, as descendants
of $\Psi_k^+$, only the left fermionic components are non-trivial in the
original Verma module according to \eq{\ref{eq:van3}},
and there exists a left fermionic component for all
solutions. Therefore, we obtain $+1$ charged singular vectors at
the levels $k+r_n^+\tilde{s}_n^+$ with $r_n^+=j+k+(n+1)m$ and
$\tilde{s}^+_n=n$ for $n\in\bbbz\setminus\{-1,0\}$. 
Once again we do not obtain more singular vectors in the 
charged sector by analysing 
more descendant vectors of $\Psi^{+}_{r_n^+,s_n^+}$, 
since the given straight line
intersects the $r$-axis at an integer point. Already at this stage, 
we ought to
mention, that by analysing charged descendant singular operators
of $\Psi^{+}_{r_n^+,s_n^+}$, we find constraints on the 
embedded Verma modules built on top of $\Psi^{+}_{r_n^+,s_n^+}$, since
there are no charge $2$ singular vectors in the 
original Verma module. For $\Psi^{+}_{r_n^+,s_n^+}$ we easily find
these constraints at descendant relative levels $k+nm$
for $n\in\bbbn$ and $-j-(n+1)m$ for $n\in -\bbbn \setminus \{ -1\}$,
simply by computing 
the new factors $k^a$ and $k^b$.
It is important to realise that these are the
descendant charged singular operators at lowest 
levels\footnote{Is is also important to note that the first
uncharged singular vector is neither left nor right fermionic. Therefore,
the charged singular vector given by the new factor
$k^a$ or $k^b$ defines the lowest
level operator annihilating $\Psi^{+}_{r_n^+,s_n^+}$.}, 
exactly as they appear in the
fermionic partition function of the determinant\cite{adrian}. 
All other singular vectors in the kernel of $\Psi_{k}^{+}$
descend from these operators, so that for our later
considerations in addition to $\Psi_{k}^{+}$ we only need to be concerned
about possible subsingular vectors in the kernel in order to
generate the whole kernel.
Similar results are valid for 
$\Psi_j^-$. 
\btab{|c|c|c|c|} 
\hline 
\ty{vector} & \ty{parameters ($\tilde{s}$ is always $=n$)}  
& \ty{levels} & \ty{kernel generator}
\\
\hline
\ty{$\Psi_{r_{-1},s_{-1}}$} & \ty{$r_{-1}= j+k-m$} 
& \ty{$m-j-k$} & \ty{-}
\\
\hline
\ty{$\Psi^{\lhd}_{r_n,s_n}$,  $\Psi^{\rhd}_{r_n,s_n}$}
& \ty{$r_n= j+k+nm$, $n\in \bbbn$}
& \ty{$n(nm+j+k)$} & \ty{cf. \eqs{\ref{eq:van1}-\ref{eq:van3}}}
\\
\hline
\ty{$\Psi^{\lhd}_{r_n,s_n}$,
$\Psi^{\rhd}_{r_n,s_n}$} & \ty{$r_n= j+k+nm$, $n\in -\bbbn$} 
& \ty{$|n|(|n|m-j-k)$} & \ty{cf. \eqs{\ref{eq:van1}-\ref{eq:van3}}}
\\
\hline
\ty{$\Psi^+_{r^+_n,s^+_n}$} & \ty{$r^+_n= j+k+(n+1)m$, 
$n\in \bbbn_0$} 
& \ty{$k+n\{(n+1)m+j+k\}$} & \ty{$k+nm$}
\\
\hline
\ty{$\Psi^+_{r^+_n,s^+_n}$} & \ty{$r^+_n\!=\! j\!+\!k\!+\!(n+1)\!m\!$, 
$n\in\! -\bbbn_{\setminus\{-1\}}$} 
& \ty{$k+|n|\{|n+1|m-j-k\}$} & \ty{$|n+1|m-j$}
\\
\hline
\ty{$\Psi^-_{r^-_n,s^-_n}$} & \ty{$r^-_n= j+k+(n+1)m$, 
$n\in \bbbn_0$}
& \ty{$j+n\{(n+1)m+j+k\}$} & \ty{$j+nm$}
\\
\hline
\ty{$\Psi^-_{r^-_n,s^-_n}$} & \ty{$r^-_n\!=\! j\!+\!k\!+\!(n+1)\!m\!$, 
$n\in\! -\bbbn_{\setminus\{-1\}}$} 
& \ty{$j+|n|\{|n+1|m-j-k\}$} & \ty{$|n+1|m-k$}
\\\hline
\etab{Singular vectors of the unitary 
minimal models.}\label{tab:discrete}

We have now shown that the singular vectors given in Fig.
\ref{pic:discrser} are necessarily contained in the embedding
structure at the levels given in Tab. \ref{tab:discrete}.
In order to verify the embedding homomorphisms drawn in Fig. 
\ref{pic:discrser}, we simply need to descend down from
the vectors given in Tab. \ref{tab:discrete}. For the uncharged
sector we explained this briefly above. Therefore, the only cases
left are the lines 
connecting charged and uncharged sectors.
If we find a match in level for such a connexion and if the vectors
are not annihilated by fermionic conditions, then 
due to \eqs{\ref{eq:van1}-\ref{eq:van2}}
we know
that the $-1$ charged singular vectors can only connect to 
left fermionic uncharged singular vectors, whilst the 
$+1$ charged singular vectors connect to 
right fermionic uncharged singular vectors
and vice versa.
We are now left with proving that the levels match in the way indicated
in Fig. \ref{pic:discrser}. We shall show this for lines from right
fermionic uncharged
singular vectors to $+1$ charged singular vectors only.
The determinant parameters for modules embedded on top of
$\Psi^{\rhd}_{r_n,s_n}$ are $k^{a}_{r_n,s_n}=k+nm$ which for $n\in
\bbbn$ implies a charge $+1$ singular vector. As $\Psi^{\rhd}_{r_n,s_n}$
is right fermionic, the descendant singular vector is non-trivial
in the original Verma module at level $n(j+k+nm)+k+nm$
which matches with the level of $\Psi^+_{r^+_n,s^+_n}$. 
In the case of negative $n$, $k^{b}_{r_n,s_n}=-j-nm$ leads to
the vector $\Psi^+_{r^+_{n-1},s^+_{n-1}}$.
This completes the proof that the embedding diagram in Fig.
\ref{pic:discrser} is at least
a subset of the embedding diagram for the discrete
series of unitary representations.
That it is also sufficient and that there are no 
subsingular vectors shall be shown in the following sections.   


\section{The representations ${\cal Q}^{m\nmid j+k}_{m,1}$}

In order to prove the embedding diagrams of the previous section we first
need to generalise theorem \ref{th:disrcser}
to a larger class of modules for which the same
embedding diagram is valid. We consider a particular class of representations,
again given by $\tilde{t}=\frac{1}{m}$ with $m\in \bbbn\setminus \{1\}$,
$h=\frac{jk-\frac{1}{4}}{m}$ and $q=\frac{j-k}{m}$ with
$k,j\in\bbbn_{\frac{1}{2}}$, and $m$ not dividing $j+k$, in symbols:
$m\nmid j+k$.
Obviously, we also find $a=\frac{j+k}{m}$, $k^a=k>0$ and $k^b=-j<0$.
The discrete 
series of unitary representations are certainly of this type but
the restriction $j,k,j+k\leq m-1$ does no longer apply. We shall call these
highest weight representations ${\cal Q}^{m\nmid j+k}_{m,1}$
with $m\in\bbbn\setminus\{1\}$, 
following conventions\footnote{$(m,1)$ refers to $\tilde{t}=\frac{1}{m}$.
Similarly, $(m,m^{\prime})$ denotes $\tilde{t}=\frac{m^{\prime}}{m}$.}
of Ref. \icite{wolfgmatth}.
It is rather simple to show that the same analysis
as performed in the previous section holds. We therefore ultimately 
come back to
the same embedding patterns. The only difference is that
the lowest level uncharged singular vector is not at level
$m-j-k$ but at level $(\beta +1)[(\beta+1) m -j-k]$ 
where $\beta$ represents the integer
value of $\frac{j+k}{m}$. 
For uncharged singular vectors with negative index $n$ 
we have to replace $r_n$ and $\tilde{s}_n$ by
$r_n=(n-\beta)m+j+k$ and $\tilde{s}_n=n-\beta$, and similarly for 
charged singular vectors.
\vbox{
\bpic{200}{180}{0}{10} \label{pic:rat}
\put(0,195){\ch{q:}}
\put(100,195){\ch{0}}
\put(50,195){\ch{-1}}
\put(150,195){\ch{+1}}
\put(98,180){\framebox(4,3){}}
\put(100,180){\line(0,-1){30}}
\put(100,150){\circle*{3}}
\put(153,148){\chtl{(\beta +1)[(\beta+1) m\!-\!j\!-\!k]}}
\put(100,150){\line(0,-1){30}}
\put(97,120){\circle*{3}}
\put(97,120){\makebox(0,0){\lch}}
\put(103,120){\circle*{3}}
\put(103,120){\makebox(0,0){\rch}}
\put(153,118){\chtl{m\!+\!j\!+\!k}}
\put(97,120){\line(0,-1){30}}
\put(103,120){\line(0,-1){30}}
\put(97,90){\circle*{3}}
\put(103,90){\circle*{3}}
\put(97,90){\makebox(0,0){\lch}}
\put(103,90){\makebox(0,0){\rch}}
\put(153,88){\chtl{(\beta +2)[(\beta+2)m\!-\!j\!-\!k]}}
\put(97,90){\line(0,-1){30}}
\put(103,90){\line(0,-1){30}}
\put(97,60){\circle*{3}}
\put(103,60){\circle*{3}}
\put(97,60){\makebox(0,0){\lch}}
\put(103,60){\makebox(0,0){\rch}}
\put(153,58){\chtl{2(2m\!+\!j\!+\!k)}}
\put(97,60){\line(0,-1){30}}
\put(103,60){\line(0,-1){30}}
\put(97,30){\circle*{3}}
\put(103,30){\circle*{3}}
\put(97,30){\makebox(0,0){\lch}}
\put(103,30){\makebox(0,0){\rch}}
\put(106,28){\chtl{(\beta +3)[(\beta+3)m\!-\!j\!-\!k]}}
\put(50,160){\circle*{3}}
\put(47,162){\chtr{j}}
\put(150,160){\circle*{3}}
\put(153,162){\chtl{k}}
\put(50,159){\line(0,-1){28}}
\put(150,159){\line(0,-1){28}}
\put(50,130){\circle{3}}
\put(47,132){\chtr{j+(\beta +2)[(\beta+1)m-j-k]}}
\put(150,130){\circle{3}}
\put(153,132){\chtl{k+(\beta +2)[(\beta+1)m-j-k]}}
\put(50,129){\line(0,-1){28}}
\put(150,129){\line(0,-1){28}}
\put(50,100){\circle{3}}
\put(47,102){\chtr{2m+2j+k}}
\put(150,100){\circle{3}}
\put(153,102){\chtl{2m+j+2k}}
\put(50,99){\line(0,-1){28}}
\put(150,99){\line(0,-1){28}}
\put(50,70){\circle{3}}
\put(47,72){\chtr{j+(\beta +3)[(\beta+2)m-j-k]}}
\put(150,70){\circle{3}}
\put(153,72){\chtl{k+(\beta +3)[(\beta+2)m-j-k]}}
\put(50,69){\line(0,-1){28}}
\put(150,69){\line(0,-1){28}}
\put(50,40){\circle{3}}
\put(47,42){\chtr{6m+3j+2k}}
\put(150,40){\circle{3}}
\put(153,42){\chtl{6m+2j+3k}}
\put(100,180){\line(5,-2){49}}
\put(100,180){\line(-5,-2){49}}
\put(100,150){\line(5,-2){49}}
\put(100,150){\line(-5,-2){49}}
\put(103,118){\line(5,-2){46}}
\put(97,118){\line(-5,-2){46}}
\put(103,88){\line(5,-2){46}}
\put(97,88){\line(-5,-2){46}}
\put(103,58){\line(5,-2){46}}
\put(97,58){\line(-5,-2){46}}
\put(103,122){\line(5,4){46}}
\put(97,122){\line(-5,4){46}}
\put(103,92){\line(5,4){46}}
\put(97,92){\line(-5,4){46}}
\put(103,62){\line(5,4){46}}
\put(97,62){\line(-5,4){46}}
\put(103,32){\line(5,4){46}}
\put(97,32){\line(-5,4){46}}
\put(97,20){\makebox(0,0){$\vdots$}}
\put(103,20){\makebox(0,0){$\vdots$}}
\put(50,30){\makebox(0,0){$\vdots$}}
\put(150,30){\makebox(0,0){$\vdots$}}
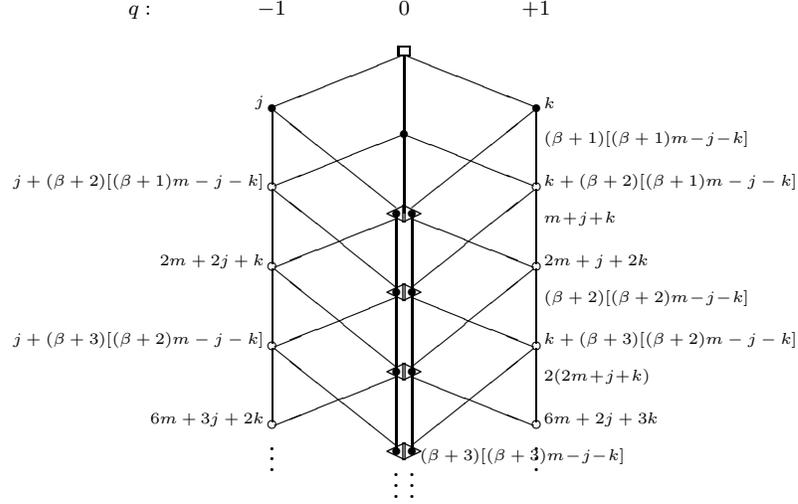
\epic{Embedding diagrams for representations 
of type ${\cal Q}^{m\nmid j+k}_{m,1}$.}}

In fact, the embedding diagrams shown in Fig. \ref{pic:rat} are valid for 
a much larger class of representations\cite{wolfgmatth,gerpet}: 
for all the representations ${\cal Q}^{m\nmid j+k}_{m,m^\prime}$
with coprime $m,m^\prime$, $k^a=k>0$, $k^b=-j<0$ and 
$m,m^\prime\in\bbbn$, $k,j\in\bbbn_{\frac{1}{2}}$,
and $m$ not dividing $j+k$,
as described in Ref. \icite{wolfgmatth}. This is difficult to understand
by just descending down along vanishing curves of the determinant. 
{\it A priori} one would believe, that the charged singular vectors
do not join the uncharged singular vectors any more, what
clearly violates a generalised asymptotic {\it Ore}-condition\cite{ore}
for $N=2$.
This belief is based on the conditions of 
\eqs{\ref{eq:van1}-\ref{eq:van3}}. 
However, in this case you have to apply two of
such conditions successively. The first condition annihilates a submodule
assumed to contain the operator connecting to the uncharged
sector. But even this connexion is subject to one of the vanishing
conditions. Therefore, the removed kernel does not contain
the connexion back to the uncharged sector. For the purpose of this paper,
we do not need to be concerned by these generalisations though,
but our proof would equally well hold for these cases.


\section{Character formulae}

Eholzer and Gaberdiel showed in Ref. \icite{wolfgmatth} 
that the embedding diagram of Fig.
\ref{pic:rat} reveals the correct vacuum character formula. 
In the same manner, we shall briefly
explain how to deduce generally characters out of the embedding
diagrams until we reach the familiar character formulae.
Finally, this will also complete the proof for
the embedding diagram of Fig. \ref{pic:rat}. 
We define the partition function 
over a $(L_0,T_0)$-graded module ${\cal V}_{h,q}$ 
as the formal power series $P_{\cal V}=x^{-h}y^{-q}
\trmod{{\cal V}}{x^{L_0} y^{T_0}}$.
We first write down all the partition functions for the modules
built on top of the singular vectors\footnote{We
will not need partition functions for $\Psi^{\lhd}_{r_n,s_n}$
and  $\Psi^{\rhd}_{r_n,s_n}$ but for unconstrained
singular vectors at the same levels which we denote by $\Psi_{r_n,s_n}$.}
given in Tab. \ref{tab:discrete}.
For convenience, we label modules ${\cal V}$ of type
${\cal Q}^{m\nmid j+k}_{m,1}$ simply by their parameters $j,k$ and $m$
as ${\cal V}_{j,k,m}$ and partition functions of a module built on top
of a singular vector $\Psi$ simply by $P_{\Psi}$.
These partition functions were first given by Boucher,
Friedan and Kent\cite{adrian}.
\bea
P_{{\cal V}_{j,k,m}} &=& \prod_{l=1}^{\infty} 
\frac{(1+x^{l-\frac{1}{2}}y)(1+x^{l-\frac{1}{2}}y^{-1})}{(1-x^l)^2} \com 
\label{eq:part1} \\
P_{\Psi_{r_{-n},s_{-n}}} &=& x^{n(nm-j-k)} 
P_{{\cal V}_{j,k,m}}  \com n\in\bbbn \com \\
P_{\Psi_{r_{n},s_{n}}} &=& x^{n(nm+j+k)} 
P_{{\cal V}_{j,k,m}}  \com n\in\bbbn \com \\
P_{\Psi^+_{r^+_n,s^+_n}} &=& \frac{x^{k+n\{(n+1)m+j+k\}}y}{1+x^{k+nm}y}
P_{{\cal V}_{j,k,m}}  \com n\in\bbbn_0 \com \\
P_{\Psi^+_{r^+_{-n-1},s^+_{-n-1}}} &=& 
\frac{x^{k+(n+1)(nm-j-k)}y}{1+x^{-j+nm}y}
P_{{\cal V}_{j,k,m}}  \com n\in\bbbn \com \\
P_{\Psi^-_{r^-_n,s^-_n}} &=& 
\frac{x^{j+n\{(n+1)m+j+k\}}y^{-1}}{1+x^{j+nm}y^{-1}}
P_{{\cal V}_{j,k,m}}  \com n\in\bbbn_0 \com \\
P_{\Psi^-_{r^-_{-n-1},s^-_{-n-1}}} &=& 
\frac{x^{j+(n+1)(nm-j-k)}y^{-1}}{1+x^{-k+nm}y^{-1}}
P_{{\cal V}_{j,k,m}}  \com n\in\bbbn \pkt \label{eq:part7}
\eea

Let us start with ${\cal V}_{j,k,m}$ being a module
of the discrete unitary series. 
Our aim is to compute the partition function for
\bea
{\cal Q}_{j,k,m}=\frac{{\cal V}_{j,k,m}}{U(\sct)\Psi_{r_{-1},s_{-1}}
+U(\sct)\Psi^+_{k}+U(\sct)\Psi^-_{j}}
\eea
and show that this reveals the correct character formula. Therefore, we first
construct the partition function for 
${\cal Q}^\prime_{j,k,m}=\frac{{\cal V}_{j,k,m}}{
U(\sct)\Psi^+_{k}+U(\sct)\Psi^-_{j}}$ which is simply
$P_{{\cal Q}^\prime_{j,k,m}}=P_{{\cal V}_{j,k,m}}-
P_{\Psi_{k}^+} -P_{\Psi_{j}^-}$ since the submodules
generated by $\Psi_{k}^+$ and  $\Psi_{j}^-$ do not 
intersect\footnote{Note that at an intersection point we would need
to find a common singular vector.}. 
${\cal Q}^\prime_{j,k,m}$ contains
${\cal Q}_{j,k,m}$ and they are related by
${\cal Q}_{j,k,m}=\frac{{\cal Q}^\prime_{j,k,m}}{U(\sct)\Psi_{r_{-1},s_{-1}}}$.
Complications arise due to the fact that $U(\sct)\Psi_{r_{-1},s_{-1}}$
intersects with the submodules generated by $\Psi_{k}^+$ and
$\Psi_{j}^-$. 
If we simply subtract $P_{\Psi_{r_{-1},s_{-1}}}$ 
from $P_{{\cal Q}^\prime_{j,k,m}}$, then
the intersection module generated by $\Psi_{r_{1},s_{1}}$, 
$\Psi^{+}_{r^+_{-2},s^+_{-2}}$ and $\Psi^{-}_{r^-_{-2},s^-_{-2}}$
is subtracted already twice, and thus needs to be added back in.
It is important to note that $\Psi_{r_{1},s_{1}}$ is the lowest
level uncharged descendant vector of $\Psi_{r_{-1},s_{-1}}$ and
is therefore a complete module (neither left fermionic nor
right fermionic). By adding the partition functions of these
modules back in, we again obtain an intersection module 
generated by  $\Psi_{r_{-2},s_{-2}}$, 
$\Psi^{+}_{r^+_{1},s^+_{1}}$ and $\Psi^{-}_{r^-_{1},s^-_{1}}$
which this time
has not been subtracted overall. Proceeding in this manner
leads to an alternating pattern of subtracting and adding partition
functions of the type of \eqs{\ref{eq:part1}-\ref{eq:part7}}.
This pattern coincides with the pattern given by Dobrev\cite{dobrev},
even though he started from the wrong embedding diagrams, neglecting
the degeneration of uncharged singular vectors in left and right fermionic
uncharged singular vectors\footnote{Note that the signs in the uncharged
sector of Fig. \ref{fig:pm} refer to complete uncharged Verma modules.}.

\vbox{
\bpic{200}{180}{0}{10} \label{fig:pm}
\put(0,195){\ch{q:}}
\put(100,195){\ch{0}}
\put(50,195){\ch{-1}}
\put(150,195){\ch{+1}}
\put(98,180){\framebox(4,3){}}
\put(100,180){\line(0,-1){30}}
\put(100,150){\circle*{3}}
\put(102,152){\chtl{-}}
\put(100,150){\line(0,-1){30}}
\put(97,120){\circle*{3}}
\put(97,120){\makebox(0,0){\lch}}
\put(103,120){\circle*{3}}
\put(103,120){\makebox(0,0){\rch}}
\put(106,122){\chtl{+}}
\put(97,120){\line(0,-1){30}}
\put(103,120){\line(0,-1){30}}
\put(97,90){\circle*{3}}
\put(103,90){\circle*{3}}
\put(97,90){\makebox(0,0){\lch}}
\put(103,90){\makebox(0,0){\rch}}
\put(106,92){\chtl{-}}
\put(97,90){\line(0,-1){30}}
\put(103,90){\line(0,-1){30}}
\put(97,60){\circle*{3}}
\put(103,60){\circle*{3}}
\put(97,60){\makebox(0,0){\lch}}
\put(103,60){\makebox(0,0){\rch}}
\put(106,62){\chtl{+}}
\put(97,60){\line(0,-1){30}}
\put(103,60){\line(0,-1){30}}
\put(97,30){\circle*{3}}
\put(103,30){\circle*{3}}
\put(97,30){\makebox(0,0){\lch}}
\put(103,30){\makebox(0,0){\rch}}
\put(106,32){\chtl{-}}
\put(50,160){\circle*{3}}
\put(47,162){\chtr{-}}
\put(150,160){\circle*{3}}
\put(153,162){\chtl{-}}
\put(50,159){\line(0,-1){28}}
\put(150,159){\line(0,-1){28}}
\put(50,130){\circle{3}}
\put(47,132){\chtr{+}}
\put(150,130){\circle{3}}
\put(153,132){\chtl{+}}
\put(50,129){\line(0,-1){28}}
\put(150,129){\line(0,-1){28}}
\put(50,100){\circle{3}}
\put(47,102){\chtr{-}}
\put(150,100){\circle{3}}
\put(153,102){\chtl{-}}
\put(50,99){\line(0,-1){28}}
\put(150,99){\line(0,-1){28}}
\put(50,70){\circle{3}}
\put(47,72){\chtr{+}}
\put(150,70){\circle{3}}
\put(153,72){\chtl{+}}
\put(50,69){\line(0,-1){28}}
\put(150,69){\line(0,-1){28}}
\put(50,40){\circle{3}}
\put(47,42){\chtr{-}}
\put(150,40){\circle{3}}
\put(153,42){\chtl{-}}
\put(100,180){\line(5,-2){49}}
\put(100,180){\line(-5,-2){49}}
\put(100,150){\line(5,-2){49}}
\put(100,150){\line(-5,-2){49}}
\put(103,118){\line(5,-2){46}}
\put(97,118){\line(-5,-2){46}}
\put(103,88){\line(5,-2){46}}
\put(97,88){\line(-5,-2){46}}
\put(103,58){\line(5,-2){46}}
\put(97,58){\line(-5,-2){46}}
\put(103,122){\line(5,4){46}}
\put(97,122){\line(-5,4){46}}
\put(103,92){\line(5,4){46}}
\put(97,92){\line(-5,4){46}}
\put(103,62){\line(5,4){46}}
\put(97,62){\line(-5,4){46}}
\put(103,32){\line(5,4){46}}
\put(97,32){\line(-5,4){46}}
\put(97,20){\makebox(0,0){$\vdots$}}
\put(103,20){\makebox(0,0){$\vdots$}}
\put(50,30){\makebox(0,0){$\vdots$}}
\put(150,30){\makebox(0,0){$\vdots$}}
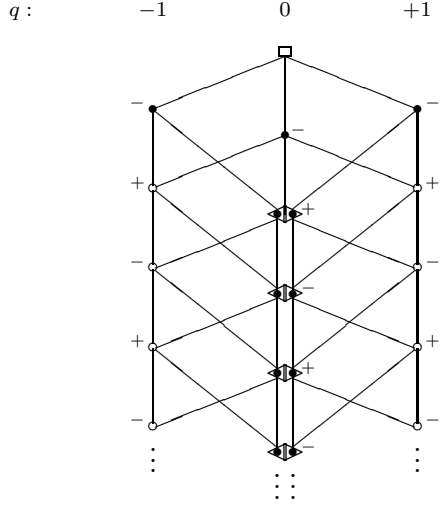
\epic{Constructing
the partition function $P_{{\cal Q}_{j,k,m}}$.}}

\noi We finally obtain the following 
partition function for ${\cal Q}_{j,k,m}$:
\bea
P_{{\cal Q}_{j,k,m}} &=& P_{{\cal V}_{j,k,m}} \big\{
1
-\sum_{n\in\bbbn_0} P_{\Psi^{+}_{r^+_{n},s^+_{n}}}
-\sum_{n\in\bbbn_0} P_{\Psi^{-}_{r^-_{n},s^-_{n}}}
-\sum_{n\in -\bbbn} P_{\Psi_{r_{n},s_{n}}} \nn \\
&& +\sum_{n\in -\bbbn\setminus\{-1\}} P_{\Psi^{+}_{r^+_{n},s^+_{n}}}
+\sum_{n\in -\bbbn\setminus\{-1\}} P_{\Psi^{-}_{r^-_{n},s^-_{n}}}
+\sum_{n\in \bbbn} P_{\Psi_{r_{n},s_{n}}} \big\} \com
\eea
\bea
&=& P_{{\cal V}_{j,k,m}} \Big\{ 1+
\sum_{n=0}^{\infty} x^{n^2m} (x^{n(j+k)}-x^{-n(j+k)}) \nn \\
&& +
\sum_{n=0}^{\infty} x^{(n+1)nm-j}y 
(\frac{x^{2(n+1)m-(n+1)j-(n+1)k}}{1+x^{(n+1)m-j}y}
-\frac{x^{(n+1)j+(n+1)k}}{1+x^{k+nm}y}) \nn \\
&& +
\sum_{n=0}^{\infty} x^{(n+1)nm-k}y^{-1} 
(\frac{x^{2(n+1)m-(n+1)j-(n+1)k}}{1+x^{(n+1)m-k}y^{-1}}
-\frac{x^{(n+1)j+(n+1)k}}{1+x^{j+nm}y^{-1}})
\Big\} \pkt \label{eq:char} 
\eea
\noi
\eq{\ref{eq:char}} coincides\footnote{Strictly speaking, 
we have to assume that for the coset construction
the $N=2$ algebra is indeed the coset and
not only a subalgebra of the coset. No
independent proof of this is known to us. 
We are grateful to Matthias Gaberdiel and
Wolfgang Eholzer for pointing this out.} with known formulae for the $N=2$
characters\cite{wolfgmatth,ahn,huitu,rava,dobrev}.
Deriving similar expressions for the representations of type 
${\cal Q}^{m\nmid j+k}_{m,1}$ is straightforward 
by adding appropriate factors of $\beta$, and is left to the
reader.
Before we finish the proof for the embedding diagrams, we 
make some important remarks:
\bth \label{th:ssvec}
$\Upsilon$ is a subsingular vector in the Verma module
${\cal V}$. If $\Upsilon$ lies in a submodule ${\cal W}$ of ${\cal V}$, 
then $\Upsilon$
is also subsingular in ${\cal W}$.
\eth
\bprf
We have nothing to show
if $\Upsilon$ is subsingular with respect to a singular vector
$\Psi\in {\cal W}$. Thus, assume that $\Psi$ lies\footnote{Instead of having
only one vector $\Psi$, $\Upsilon$ can also be subsingular with respect
to many singular and subsingular vectors. The proof still holds.}
outside ${\cal W}$. The module built on top of $\Psi$ cannot be disjoint from
${\cal W}$ otherwise $\Upsilon$ would be singular. The intersection is again
given by a submodule ${\cal Y}\subset {\cal W}$. 
The positive algebra modes either annihilate
$\Upsilon$ or map $\Upsilon$ inside ${\cal Y}$. 
Therefore $\Upsilon$ is subsingular with
respect to the singular (and subsingular) vectors generating ${\cal Y}$. 
${\cal Y}$ cannot be $\equiv {\cal W}$ because $\Upsilon$ would in this
case lie in the module generated on top of $\Psi$.
\eprf

\bth \label{th:dd}
For a representation of type ${\cal Q}^{m\nmid j+k}_{m,1}$ the uncharged 
singular vector
at lowest level $\Psi_{r_{-1},s_{-1}}$ and the
charged vectors $\Psi^{+}_{k}$ and $\Psi_{j}^{-}$ 
define as unconstrained highest weight modules
${\cal V}_{h_{r_{-1},s_{-1}}+r_{-1}\tilde{s}_{-1},q,t}$,
${\cal V}_{h^+_{k}+k,q+1,t}$ and ${\cal V}_{h^-_{j}+j,q-1,t}$
again representations
of type ${\cal Q}^{m\nmid j+k}_{m,1}$. 
\eth
\bprf
If the given model has the parameters $k^a=k$ and $k^b=-j$,
$k,j\in\bbbn_{\frac{1}{2}}$, then the factors $k_{-1}^a$ and $k_{-1}^b$
for the module ${\cal V}_{h_{r_{-1},s_{-1}}+r_{-1}\tilde{s}_{-1},q,t}$
with $r_{-1}=j+k-(\beta+1)m$, $\tilde{s}=-(\beta +1)$ are:
$k^a_{-1}=k-(\beta+1)m$, $k^b_{-1}=-j+(\beta+1)m$. Which is again
a representation of type\footnote{Note that $(\beta +1)m\geq j,k$.}
${\cal Q}^{m\nmid j+k}_{m,1}$. Similarly,
$k^a_{+}= k$, $k^{b}_{+}=-j-m$, $k^a_{-}= k+m$ and $k^{b}_{-}=-j$.
\eprf

In order to conclude 
the proof of the embedding diagrams two things need to be shown.
Firstly, there are no subsingular vectors in
the highest weight modules of type ${\cal Q}^{m\nmid j+k}_{m,1}$
and secondly, the kernels of the fermionic singular vectors 
do not contain any
subsingular vectors.
Let ${\cal V}$ be a Verma module of type 
${\cal Q}^{m\nmid j+k}_{m,1}$ and let ${\cal W}^0$,
${\cal W}^+$ and ${\cal W}^-$ denote the
submodules built on top of $\Psi_{r_{-1},s_{-1}}$,
$\Psi^+_{k}$ and $\Psi^-_{j}$ respectively.
Furthermore, ${\cal V}^{\pm}$ are the (complete)
Verma modules with $\Psi^{\pm}$ as unconstrained highest weight
vector which means ${\cal W}^{\pm}=\frac{{\cal V}^{\pm}}{\ker\{\Psi^{\pm}\}}$.
Obviously, the module ${\cal V}$ cannot contain any subsingular vectors
at levels $l$ with $l \leq min(j,k,r_{-1}\tilde{s}_{-1})$.
Suppose that ${\cal V}$ has a subsingular vector $\Upsilon$ at level 
$l > min(j,k,r_{-1}\tilde{s}_{-1})$ with charge $q$ 
and that $\Upsilon$ is not contained in 
${\cal W}^0 \cup {\cal W}^+ \cup {\cal W}^-$. In this case, the coefficient
of the partition function $P_{{\cal Q}_{j,k,m}}$ for the grade 
$(l,q)$ is too big, since we need to subtract $\Upsilon$ (and all descendants
of $\Upsilon$). However, this coefficient agrees with
the correct character formula. Thus there needs to exist another
subsingular vector $\Upsilon^\prime$ 
(possibly at different level and charge) which
compensates the subtraction of $\Upsilon$, 
i.e. $\Upsilon^\prime$ is being added rather than subtracted.
Such a subsingular vector could perhaps be found in the kernel of the
fermionic singular vectors. If there is a subsingular 
vector in the kernel, then the fermionic modules
are in fact smaller than we thought and we have originally
subtracted too much since some of the expressions of 
\eqs{\ref{eq:part1}-\ref{eq:part7}} 
have in that case too big coefficients.
Nevertheless, $\Upsilon^\prime$ can certainly
not be outside ${\cal W}^0 \cup {\cal W}^+ \cup {\cal W}^-$, because if
it was, then it would be subtracted rather than added. 
Therefore we have either
$\Upsilon^\prime \in {\cal W}^0 \cup {\cal W}^+ \cup {\cal W}^-$,
$\Upsilon^\prime \in {\cal V}^+$ or $\Upsilon^\prime \in {\cal V}^-$.
Note that the level of $\Upsilon$ cannot be smaller than the
level of $\Upsilon^\prime$. 
According to theorem \ref{th:ssvec}, $\Upsilon^\prime$ is then even 
subsingular in 
at least one of the (complete) Verma modules
${\cal W}^0$, ${\cal V}^+$ or ${\cal V}^-$.
Theorem \ref{th:dd} tells us then, that this Verma
module is again of type ${\cal Q}^{m\nmid j+k}_{m,1}$ and therefore the same
procedure starts all over again. By induction, we can 
reach any arbitrary level 
in the embedding patterns free of subsingular vectors. 
This completes the proof.
\eprf

\noi We summarise our results in the following theorem:
\bth
The embedding diagrams of the unitary $N=2$ minimal models
and modules of type ${\cal Q}^{m\nmid j+k}_{m,1}$
are given by Fig. \ref{pic:discrser} and Fig. \ref{pic:rat}.
There are no subsingular vectors in these representations.
Furthermore, the singular vectors $\Psi^{\pm}_{r^{\pm}_n,s^{\pm}_{n}}$
have constraints generated by the singular operators
given in Tab. \ref{tab:discrete} and their kernels do not contain 
any subsingular vectors.
\eth

Just recently subsingular vectors were discovered for
some highest weight representations of the $\sct$ algebra
by Gato-Rivera and 
Rosado\cite{beatriz1,beatriz2}. As we can easily verify, in the case
of the unitary minimal models these subsingular vectors turn into
ordinary singular vectors already contained in Fig. \ref{pic:discrser}.
It is also easy to verify that 
the subsingular vectors given in Ref. \icite{paper7} are not 
representations of
type ${\cal Q}^{m\nmid j+k}_{m,1}$.

\section{Conclusion}
In this paper we have presented the embedding structure for
singular vectors of the unitary minimal models and for representations 
of type ${\cal Q}^{m\nmid j+k}_{m,1}$. These embedding
patterns show a degeneration of uncharged singular vectors.
We have therefore corrected 
previous publications in this area by different 
authors\cite{dobrev,kiritsis2,matsuo}.
The character expressions derived from our embedding patterns
coincide with the correct character formulae for these representations.
Surprisingly enough, the character formulae also agrees
with the ones derived earlier by the above authors
using wrong embedding structures. We finally gave a proof
for the embedding diagram using the character formula.
This shows, that the $N=2$ unitary minimal models do not contain
any subsingular vectors. Nevertheless, there are other 
representations of the $N=2$ superconformal algebra which
do contain subsingular vectors as recently shown by
Gato-Rivera and Rosado\cite{beatriz1}. The fact that just
the unitary representations are not concerned by subsingular vectors
raises the question whether this holds for a much larger
class of algebras.  

\appendix

\acknowledgements
\noi
I am extremely grateful to Adrian Kent, Matthias 
Gaberdiel and Wolfgang Eholzer 
for many illuminating discussions on various aspects
of this work.
I am also indebted to Beatriz Gato-Rivera
for many discussions about subsingular vectors.
I would like to thank Peter Bowcock and G\'erard Watts
for numerous helpful comments. 
Finally, I am grateful to my wife Val\'erie for her kind help in
linguistic matters.
This work has been supported by a DAAD fellowship and in part by 
NSF grant PHY-92-18167.


\newcommand{\tit}[1]{{\it #1}}

\end{document}